\documentstyle[twoside,epsf]{article}

\catcode`\@=11
\long\def\@makefntext#1{
\protect\noindent \hbox to 3.2pt {\hskip-.9pt  
$^{{\eightrm\@thefnmark}}$\hfil}#1\hfill}               

\def\@makefnmark{\hbox to 0pt{$^{\@thefnmark}$\hss}}    
        
\def\ps@myheadings{\let\@mkboth\@gobbletwo
\def\@oddhead{\hbox{}
\rightmark\hfil\eightrm\thepage}   
\def\@oddfoot{}\def\@evenhead{\eightrm\thepage\hfil
\leftmark\hbox{}}\def\@evenfoot{}
\def\sectionmark##1{}\def\subsectionmark##1{}}

\oddsidemargin=\evensidemargin
\newcounter{sectionc}\newcounter{subsectionc}\newcounter{subsubsectionc}
\renewcommand{\section}[1] {\vspace{12pt}\addtocounter{sectionc}{1} 
\setcounter{subsectionc}{0}\setcounter{subsubsectionc}{0}\noindent 
        {\tenbf\thesectionc. #1}\par\vspace{5pt}}
\renewcommand{\subsection}[1] {\vspace{12pt}\addtocounter{subsectionc}{1} 
\setcounter{subsubsectionc}{0}\noindent 
{\bf\thesectionc.\thesubsectionc. {\kern1pt \bfit #1}}\par\vspace{5pt}}
\renewcommand{\subsubsection}[1] {\vspace{12pt}\addtocounter{subsubsectionc}{1}
        \noindent{\tenrm\thesectionc.\thesubsectionc.\thesubsubsectionc.
        {\kern1pt \tenit #1}}\par\vspace{5pt}}

\newcounter{appendixc}
\newcounter{subappendixc}[appendixc]
\newcounter{subsubappendixc}[subappendixc]
\renewcommand{\thesubappendixc}{\Alph{appendixc}.\arabic{subappendixc}}
\renewcommand{\thesubsubappendixc}
        {\Alph{appendixc}.\arabic{subappendixc}.\arabic{subsubappendixc}}

\renewcommand{\appendix}[1] {\vspace{12pt}
        \refstepcounter{appendixc}
        \setcounter{figure}{0}
        \setcounter{table}{0}
        \setcounter{lemma}{0}
        \setcounter{theorem}{0}
        \setcounter{corollary}{0}
        \setcounter{definition}{0}
        \setcounter{equation}{0}
        \renewcommand{\thefigure}{\Alph{appendixc}.\arabic{figure}}
        \renewcommand{\thetable}{\Alph{appendixc}.\arabic{table}}
        \renewcommand{\theappendixc}{\Alph{appendixc}}
        \renewcommand{\thelemma}{\Alph{appendixc}.\arabic{lemma}}
        \renewcommand{\thetheorem}{\Alph{appendixc}.\arabic{theorem}}
        \renewcommand{\thedefinition}{\Alph{appendixc}.\arabic{definition}}
        \renewcommand{\thecorollary}{\Alph{appendixc}.\arabic{corollary}}
        \renewcommand{\theequation}{\Alph{appendixc}.\arabic{equation}}
        \noindent{\tenbf Appendix \theappendixc #1}\par\vspace{5pt}}
\newcommand{\subappendix}[1] {\vspace{12pt}
        \refstepcounter{subappendixc}
        \noindent{\bf Appendix \thesubappendixc. {\kern1pt \bfit #1}}
        \par\vspace{5pt}}
\newcommand{\subsubappendix}[1] {\vspace{12pt}
        \refstepcounter{subsubappendixc}
        \noindent{\rm Appendix \thesubsubappendixc. {\kern1pt \tenit #1}}
        \par\vspace{5pt}}

\newcommand{\textlineskip}{\baselineskip=13pt}
\newcommand{\smalllineskip}{\baselineskip=10pt}

\newcommand{\copyrightheading}[1]
        {\vspace*{-2.5cm}\smalllineskip{\flushleft
        {\footnotesize Quantum Information and Computation, Vol.~1, No.~0 (2001) 000--000 #1}\\
        {\footnotesize \copyright\kern2pt Rinton Press}\\
         }}

\newcommand{\publisher}[2]{{\begin{center}\footnotesize\smalllineskip 
        Received #1\\
        Revised #2
        \end{center}
        }}

\def\abstracts#1#2#3{{
        \centering{\begin{minipage}{4.5in}\footnotesize\baselineskip=10pt
        \parindent=0pt #1\par 
        \parindent=15pt #2\par
        \parindent=15pt #3
        \end{minipage}}\par}} 

\def\keywords#1{{
        \centering{\begin{minipage}{4.5in}\footnotesize\baselineskip=10pt
        {\footnotesize\it Keywords}\/: #1
         \end{minipage}}\par}}
\def\communicate#1{{
        \centering{\begin{minipage}{4.5in}\footnotesize\baselineskip=10pt
        {\footnotesize\it Communicated by}\/: #1
         \end{minipage}}\par}}

\renewenvironment{thebibliography}[1]
        {\frenchspacing
         \ninerm\baselineskip=11pt
         \begin{list}{\arabic{enumi}.}
        {\usecounter{enumi}\setlength{\parsep}{0pt}     
         \setlength{\leftmargin 12.7pt}{\rightmargin 0pt}
         \setlength{\itemsep}{0pt} \settowidth
        {\labelwidth}{#1.}\sloppy}}{\end{list}}

\newcounter{itemlistc}
\newcounter{romanlistc}
\newcounter{alphlistc}
\newcounter{arabiclistc}

\newcommand{\fcaption}[1]{
        \refstepcounter{figure}
        \setbox\@tempboxa = \hbox{\footnotesize Fig.~\thefigure. #1}
        \ifdim \wd\@tempboxa > 5in
           {\begin{center}
        \parbox{5in}{\footnotesize\smalllineskip Fig.~\thefigure. #1}
            \end{center}}
        \else
             {\begin{center}
             {\footnotesize Fig.~\thefigure. #1}
              \end{center}}
        \fi}

\newcommand{\tcaption}[1]{
        \refstepcounter{table}
        \setbox\@tempboxa = \hbox{\footnotesize Table~\thetable. #1}
        \ifdim \wd\@tempboxa > 5in
           {\begin{center}
        \parbox{5in}{\footnotesize\smalllineskip Table~\thetable. #1}
            \end{center}}
        \else
             {\begin{center}
             {\footnotesize Table~\thetable. #1}
              \end{center}}
        \fi}

\def\pmb#1{\setbox0=\hbox{#1}
        \kern-.025em\copy0\kern-\wd0
        \kern.05em\copy0\kern-\wd0
        \kern-.025em\raise.0433em\box0}

\def\fnt#1#2{\footnotetext{\kern-.3em
        {$^{\mbox{\scriptsize #1}}$}{#2}}}

\def\fpage#1{\begingroup
\voffset=.3in
\thispagestyle{empty}\begin{table}[b]\centerline{\footnotesize #1}
        \end{table}\endgroup}

\def\runninghead#1#2{\pagestyle{myheadings}
\markboth{{\protect\footnotesize\it{\quad #1}}\hfill}
{\hfill{\protect\footnotesize\it{#2\quad}}}}
\font\tenrm=cmr10
\font\tenit=cmti10 
\font\tenbf=cmbx10
\font\bfit=cmbxti10 at 10pt
\font\ninerm=cmr9

\font\eightrm=cmr8

\def\FigName{figure}%
\newbox\captionbox
\long\def\@makecaption#1#2{%
  \ifx\FigName\@captype
    \vskip\abovecaptionskip
    \setbox\tempbox\hbox{{\figurecaptionfont #1\hskip1em #2}}
        \ifdim\wd\tempbox< 28pc
        \centerline{\box\tempbox}
        \else
        {\figurecaptionfont #1\hskip1em #2\par}
\fi\else
        \setbox\tempbox\hbox{{\tablecaptionfont #1\hskip1em #2}}
        \ifdim\wd\tempbox< 28pc 
        \centerline{\box\tempbox}
        \else
        {\tablecaptionfont #1\hskip1em #2\par}%
        \fi   
 \vskip\belowcaptionskip
 \fi}
\InputIfFileExists{psfig.sty}
{\typeout{^^Jpsfig.sty inputed...ok}}{\typeout{^^JWarning: psfig.sty could be be found.^^J}}
\InputIfFileExists{epsfsafe.tex}
{\typeout{^^Jepsfsafe.tex inputed...ok}}
                        {\typeout{^^JWarning: epsfsafe.tex could not be found.^^J}}
\InputIfFileExists{epsfig.sty}
{\typeout{^^Jepsfig.sty inputed...ok}}{\typeout{^^JWarning: epsfig.sty could not be found.^^J}}
\InputIfFileExists{epsf.sty}
{\typeout{^^Jepsf.sty inputed...ok}}{\typeout{^^JWarning: epsf.sty could not be found.^^J}}%
\def\fps@figure{tbp}
\def\ftype@figure{1}
\def\ext@figure{lof}
\def\fnum@figure{Fig.\ \thefigure}
\def\qed{\hbox{${\vcenter{\vbox{                  
   \hrule height 0.4pt\hbox{\vrule width 0.4pt height 6pt
   \kern5pt\vrule width 0.4pt}\hrule height 0.4pt}}}$}}

\begin{document}
\setlength{\textheight}{8.0truein}    

\runninghead{Speed-up and entanglement in quantum searching} 
            {S.\ L.\ Braunstein and A.\ K.\ Pati}

\normalsize\textlineskip
\thispagestyle{empty}
\setcounter{page}{1}

\copyrightheading{}     

\vspace*{0.88truein}

\fpage{1}
\centerline{\bf
SPEED-UP AND ENTANGLEMENT IN QUANTUM SEARCHING}
\vspace*{0.035truein}

\centerline{\footnotesize 
Samuel L.\ Braunstein and Arun K.\ Pati$^{*}$ }
\vspace*{0.015truein}
\centerline{\footnotesize\it Informatics, Bangor University, Bangor, 
LL57 1UT, UK}
\vspace*{0.015truein}
\centerline{\footnotesize and}
\centerline{\footnotesize\it $^{*}$Institute of Physics, Sainik School Post}
\baselineskip=10pt
\centerline{\footnotesize\it Bhubaneswar-751005, Orissa, India.}

\publisher{(received date)}{(revised date)}

\def\openone{\leavevmode\hbox{\small1\kern-3.8pt\normalsize1}}%

\vspace*{0.21truein}
\abstracts{
We investigate the issue of speed-up and the necessity of entanglement in
Grover's quantum search algorithm. We find that in a pure state 
implementation of Grover's algorithm entanglement is present even though 
the initial and target states are product states. In pseudo-pure state
implementations, the separability of the states involved defines an
entanglement boundary in terms of a bound on the purity parameter. Using 
this bound we investigate the necessity of entanglement in quantum 
searching for these pseudo-pure state implementations. If every active 
molecule involved in the ensemble is `charged for' then in existing
machines speed-up without entanglement is not possible.}{}{}

\vspace*{10pt}
\vspace*{3pt}
\communicate{to be filled by the Editorial}

\vspace*{1pt}\textlineskip      

\vspace*{-0.5pt}
\noindent





%

\section{Introduction}

It was Feynman's insight to realize that quantum systems cannot be 
efficiently simulated on conventional classical computers \cite{rf}. 
Subsequently, Deustch suggested that if one could build computers 
exploiting the principles of quantum theory one might be able to speed 
up the computation process compared to classical approaches \cite{dd}. 
Quantum computers aim to make use of quantum interference and 
entanglement between different parts of a bulk quantum system to give 
such an essential difference \cite{rj}.

In recent years an important question has been: What will it take for 
quantum computers to surpass their conventional classical counterparts 
in speed and efficiency? The answer consists of algorithms and their 
efficient implementation. The first promising algorithms for quantum 
computers were discovered by Deutsch and Jozsa for function testing 
\cite{dj}, by Shor for factoring \cite{ps} and by Grover for searching 
\cite{lg}. While the efficiency of these algorithms is today well 
established, the conditions for achieving quantum efficiencies have been 
the subject of recent controversy \cite{PT}. Indeed, experimental 
implementations of Grover's algorithm on pseudo-pure state machines 
using liquid-state Nuclear Magnetic Resonance (NMR) \cite{cfh,gc} have 
been claimed to already achieve quantum efficiencies 
\cite{ch,jmh,vetal0,vetal} in spite of their apparent inability to have 
produced entanglement to date \cite{sam}.
Similarly, the Deutsch-Jozsa algorithm has been implemented on NMR quantum 
computers \cite{lbf,dak}. 

The aim of this paper is to argue that the original version of Grover's
algorithm \cite{lg} on multiple qubits necessarily involves quantum 
entanglement, even though the initial and target states are product states. 
Further, by counting each active molecule as contributing to the 
computational resources for pseudo-pure state machines, we show in a
non-asymptotic analysis that not only is entanglement necessary to achieve 
a speed-up in quantum searching, but it must be present throughout the 
computation. If one uses a different resource counting method there may 
be speed-up without entanglement. Thus, we have resolved the role
played by entanglement in quantum searching on one whole class of quantum 
computers. This non-asymptotic result unequivocally proves that a class of  
liquid-state NMR machines \cite{ch,jmh,vetal0,vetal} cannot be used for 
performing faster-than-classical quantum computation.

\section{Efficiency in Quantum Computation}

Typically, efficiency is quantified by relative `speed' or how the number 
of steps needed to complete an algorithm {\it scales\/} with the size of 
the `input' the algorithm is fed. Two ubiquitous `exponential' problems 
are searching and factoring: All known algorithms for solving them on 
conventional computers scale roughly exponentially with input size 
(e.g., the length of the list to be searched or size of the number 
to be factored). Discoveries of `fast' quantum algorithms \cite{dj,ps,lg} 
set new bounds on computational goals and standards.

In order to determine the efficiency of an algorithm on a quantum computer, 
the conventional measure of `speed' must first be re-evaluated. Clearly, 
for the scaling behavior to be a sensible measure of efficiency that may 
be used to compare the performance of very different kinds of computers, 
there must be no `hidden costs' that grow in an unreasonable manner 
(i.e., faster than the scaling itself). For example, any increase in the 
size of the computer itself, or number of resources it utilizes relative 
to the input, should not exceed the scaling of the number of steps.

Pseudo-pure state quantum computers are welcome candidates for such 
re-valuations. Indeed, such machines have been implemented using
liquid-state NMR and have been proposed for a wide variety of quantum 
algorithms \cite{cfh,gc,lbf,dak,nkl} and their efficiency has already been 
determined for asymptotically large systems (i.e., many qubits) for Shor's 
factoring algorithm. It was found that, in this asymptotic limit, an 
absence of entanglement would lead to an exponential decrease in the 
probability for obtaining the correct answer \cite{lp}. Thus, if Shor's 
algorithm were to be implemented on such machines, an exponentially 
large number of resources would be required to boost this low probability. 
Because Shor's algorithm provides an exponential speed-up, it may not be 
so surprising that the weirdest features of quantum mechanics, i.e., 
entanglement, are required for its implementation.

An important caveat of this analysis stems from the extreme unlikelihood
of constructing a sufficiently large machine that could be subjected to 
such asymptotic analysis. For example, it was observed that for 
liquid-state NMR implementations the signal scales as $n/2^n$ with 
increasing numbers $n$ of qubits (i.e., active spins) \cite{ww}. 
This inherent scaling problem suggests that regardless of entanglement or 
speed-up there seems little hope of ever reaching the asymptotic regime. 

The situation is very different for Grover's search algorithm. First, 
Grover's quantum algorithm provides a much more modest quadratic (as
opposed to exponential) speed-up over any search on a conventional
computer; thus one might expect it to be more robust with respect to
a loss of entanglement. Secondly, there have now been several experiments 
demonstrating this algorithm on small (few-qubit) NMR machines and 
claiming quantum efficiencies \cite{ch,jmh,vetal0,vetal}. For these 
experiments, the asymptotic signal scaling mentioned above is not relevant, 
and the entanglement question thus becomes the primary consideration in 
assessing speed-up. 

In fact, speed-up in Grover's algorithm can be evaluated quite naturally 
in terms of query complexity. The query complexity formulation yields a
non-asymptotic result which may be applied to any size problem. We argue 
that, in pure-state as well as pseudo-pure state implementations, if only 
separable (i.e., unentangled) states are accessed then the speed-up 
predicted by Grover's algorithm fails to materialize. There exists a 
peculiar exception (though one {\it not\/} accessible to existing 
pseudo-pure state machines) for a search space of size four where 
entanglement is not necessary. Since generalizations of this interesting 
exception are as yet unknown, the heuristic claim that entanglement is 
necessary for a {\it scalable\/} quantum computation implementing Grover's 
algorithm still holds. Our conclusion is also consistent with the analysis 
of Grover's algorithm, implemented on a device exploiting only 
superposition, but at the cost of scalability due to exponentially 
growing resources \cite{lloyd}. It should be noted that in addition to the
necessary condition we analyze the sufficient condition for speed-up
of quantum searching on pseudo-pure state machines.

The outline of the paper is as follows. In section 3, we begin with a 
straight-forward demonstration that the pure-state version of Grover's 
algorithm involves entanglement during the quantum searching operation. 
In section 4, we analyze the quantum search that can be implemented on 
pseudo-pure states and derive a bound on the purity parameter for
the pseudo-pure states to be separable. In section 5, to determine the 
necessity of entanglement in the pseudo-pure state version, 
two complementary criteria are derived: one for the presence of 
entanglement (as a function of the number of qubits) and one for the query 
complexity of the algorithm, yielding a measure of speed-up relative to 
the classical algorithm. We impose these two criteria in the few-qubit 
regime, one at a time, and obtain a one-to-one relation between 
entanglement and speed-up, showing both that there can be no speed-up 
without entanglement, and, conversely, that in case a speed-up is in fact 
achieved, entanglement must be present throughout the computation. In 
section 6, we discuss how pseudo-pure states are not good candidates for 
mimicking pure-state dynamics. Finally, we discuss the implications of our 
results.

\section{Quantum searching with pure states and entanglement}

According to the standard formulation of the search problem \cite{lg}, 
we are given an unknown binary function $f(x)$, which returns $1$ for 
a unique `target' value $x=y$ and $0$ otherwise, over the domain 
$x =0,\,1,\,2,\,\ldots,\,N-1$, with $N=2^n$. Our goal is to find $y$ such 
that $f(y) =1$. In Grover's algorithm, the $N$ inputs are mapped onto 
the states of $n$ quantum bits (qubits) such as spin-$\frac{1}{2}$ 
particles. The quantum problem thus becomes one of maximizing the 
overlap between the state of these $n$ qubits and target state 
$\arrowvert y\rangle$. This is equivalent to maximizing the probability 
of obtaining the desired state upon measurement. The initial state of 
these qubits is taken to be an equal superposition of all possible bit 
strings, i.e.,
\begin{equation}
\arrowvert \Psi_0 \rangle = \frac{1}{\sqrt N} \sum_{x=0}^{N-1} 
\arrowvert x \rangle \label{initialstate} \;.
\end{equation}
The Grover operator defined as $G= -I_0\,H^{\otimes n} I_y\, H^{\otimes n}$
is used repeatedly in the algorithm, where 
$I_0=\openone - 2\arrowvert \Psi_0 \rangle \langle \Psi_0 \arrowvert$, 
$I_y=\openone - 2\arrowvert y \rangle \langle y \arrowvert$ with 
$\arrowvert y \rangle$ being the target (ideally the final) state and 
$H$ is the Hadamard transformation. Thus, the Grover operator corresponds 
to a small rotation in the two-dimensional subspace spanned by the initial 
and target states. Each such rotation requires a {\it single\/} evaluation 
of $f$. Thus, unlike a classical search, the quantum search monotonically 
rotates the state towards the target.

Let us start by considering the pure-state version of Grover's algorithm.
After $k$ iterations of the Grover operator the combined $n$ qubit 
state (\ref{initialstate}) evolves to \cite{cz}
\begin{equation}
\arrowvert \Psi_k \rangle = {\cos \theta_k \over \sqrt{N -1}} 
\sum_{x \not= y} \arrowvert x \rangle 
+ \sin \theta_k \arrowvert y \rangle \;,
\label{Psik}
\end{equation}
where $\theta_k= (2k+1)\theta_0$ and $\theta_0$ satisfies 
$\sin\theta_0 = 1/\sqrt{N}$. The search is complete when
$\theta_k\simeq \pi/2$ which takes $O(\sqrt{N})$ iterations of the
Grover operator and hence this many evaluations of the function $f$.

We will now show that although the initial and target states are product
states the intermediate states through which system evolves are entangled. 
Since these states are superpositions of product states they are expected
to be entangled. But how much entanglement is there in these intermediate 
states? This is difficult to quantify, as we do not have a proper measure 
of entanglement for quantum systems consisting of an arbitrary number of 
subsystems. However, we can consider the system as being bipartite, with 
one subsystem consisting of a single qubit and the second subsystem all 
the rest. In this way, we will be able to quantify the bipartite 
entanglement by calculating the reduced density matrix of any single 
qubit. We then ask how close this reduced state is to a maximally 
entangled qubit (using say the Hilbert-Schmidt norm criterion).

The reduced density matrix of the $\ell$th qubit is 
\begin{eqnarray}
\rho_\ell(k) &=&  {\rm tr}_{1,2,..,\ell-1,\ell+1,..n}\,
( \arrowvert \Psi_k \rangle  
\langle \Psi_k \arrowvert ) \nonumber \\
&=& a_k^2 H\arrowvert 0 \rangle \langle 0 \arrowvert H + 
b_k^2 \arrowvert j_\ell \rangle  \langle j_\ell \arrowvert 
+ {a_k b_k \over \sqrt{N }} (2 \arrowvert j_\ell \rangle  \langle  
j_\ell \arrowvert + \arrowvert {\tilde j}_\ell \rangle
\langle j_\ell \arrowvert + \arrowvert j_\ell \rangle  
\langle {\tilde j}_\ell 
\arrowvert ) \label{reducedD} \;,
\end{eqnarray}
where $a_k = \sqrt{N/(N-1)}\cos \theta_k$, 
$b_k = \sin \theta_k - \cos \theta_k/\sqrt {N-1}$ and the single bit
${\tilde j}_\ell = 1 -j_\ell$, $(j_\ell =0,1)$. Without loss
of generality we take $j_\ell =1$ and the density matrix $\rho_\ell(k)$ 
can be expressed in standard form, i.e.,
\begin{equation}
\rho_\ell(k) =  \frac{1}{2} [\openone + {\vec s}(k)\cdot {\vec\sigma} ] 
= [1 -s(k)] \frac{\openone}{2} + s(k) P \label{red},
\end{equation}
where ${\vec s}(k) \equiv {\rm tr}\,[\rho_\ell(k) {\vec\sigma} ]$,  
${\vec s}(k)\cdot {\vec s}(k) = s(k)^2\le 1$ and $P$ is a pure state 
projector. The components of the Bloch vector ${\vec s}(k)$ after 
$k$ iterations are 
\begin{eqnarray}
s_x(k)&=&\frac{N-2}{N-1}\cos^2\theta_k+{1\over\sqrt{N -1}} \sin 2\theta_k
\nonumber \\
s_y(k)&=&0 \nonumber \\
s_z(k)&=&\frac{1}{N-1}\cos^2 \theta_k- \sin^2 \theta_k  
\label{Blochvec} \;.
\end{eqnarray}

The bipartite entanglement in the pure state may be characterized by 
calculating the von Neumann entropy of this reduced state.
Using the expansion formula 
\begin{equation}
\log\Bigl( \frac{1 -s}{2} \openone + s P \Bigr) 
= \log\Bigl( \frac{1-s}{2} \Bigr) + P\, \log \Bigl(1 + \frac{2s}{1-s}\Bigr) \;,
\end{equation}
which holds for any $0\le s< 1$, the von Neumann entropy may be
calculated to be given by
\begin{eqnarray}
S[\rho_\ell(k)] &=&  - {\rm tr}\,[ \rho_\ell(k)  \log \rho_\ell(k) ]  
\nonumber\\
&=& 1- \frac{1-s(k)}{2}\log [1-s(k)] - \frac{1+s(k)}{2} \log [1+s(k)] \;.
\end{eqnarray}
The right-hand-side of this expression is independent of the choice of 
the remaining qubit $\ell$. 
Therefore, (7) holds for any one qubit versus $(n-1)$ qubit partioning.
It shows that the reduced density matrix of the
single qubit does not arise from a maximally entangled state of 
$n$ qubits, as the von Neumann entropy is not exactly unity. Since the 
reduced state of Eq.~(\ref{reducedD}) is not pure the full state must be 
entangled. To see how impure the state in Eq.~(\ref{reducedD}) is one may 
calculate the linear entropy $L(\rho)$ of it which is given by
\begin{equation}
L[\rho_\ell(k)] =  {\rm tr}\,[ \rho_\ell(k) - \rho_\ell(k)^2] =
\frac{1 -  s(k)^2}{2}\;.
\end{equation}
If the linear entropy is zero the state is pure and as it approaches
$\frac{1}{2}$ the state approaches a completely random mixture. In the 
quantum search algorithm the parameter $s(k)$ can never be zero because that 
would mean that $\cos \theta_k$ and $\sin \theta_k$ are simultaneously 
zero, which cannot be satisfied. So although the reduced density matrix of 
the qubit may lie close to the completely mixed state it can never become
the identity one. 

We now calculate the Hilbert-Schimdt norm of the difference of two density 
matrices: namely, the completely mixed one and the reduced state given by 
Eq.~(\ref{reducedD}). This will yield a measure of closeness to the
completely mixed state. This Hilbert-Schmidt distance for $k$th iteration 
during quantum search algorithm is given by
\begin{eqnarray}
d(k)^2 &=& \parallel\frac{\openone}{2} - \rho_\ell(k) 
\parallel_{HS}^2 ={\rm tr}\,[\frac{\openone}{2} - \rho_\ell(k) ]^2 \nonumber \\
&=& \frac{1}{2} - L[\rho_\ell(k)] = \frac{s(k)^2}{2} \;.
\label{eq9}
\end{eqnarray}
The distance $d(k)$ provides an idea of how the reduced state of an 
individual qubit behaves during the $k$th iteration.
It shows that the reduced density matrix of the qubit differs from a 
completely random mixture by an order of $O(s(k))$. One can see from 
Eq.~(\ref{Blochvec}) and~(\ref{eq9}) that for 
$\theta_0=\sin^{-1}(1/\sqrt N)$ and
for $\theta_k=\pi/2$ the reduced density matrix of any remaining qubit
is pure, implying that the whole state must have been non-entangled.
Thus, we see that although the initial and target states are separable, 
the intermediate states through which the system evolves are 
always entangled.

\section{Quantum searching with pseudo-pure state and entanglement}

In the following we discuss the issue of entanglement in quantum searching 
with pseudo-pure states. These states naturally arise in liquid-state NMR 
machines. Here one faces the difficulty of accessing a pure state because 
the system is in thermal equilibrium at room temperature. Instead, one 
implements Grover's algorithm on a near random ensemble of molecular spins 
in a liquid, with a small preference for the spins to point along an 
external magnetic field; the size of this preference is quantified by the 
purity parameter $\epsilon$ [typically $O(10^{-5})$] 
\cite{cfh,gc,ch,jmh,vetal0,vetal}. For a sufficiently low spin polarization 
(corresponding to a sufficiently low purity parameter), the system can be 
well-approximated by a pseudo-pure state representation described by
\begin{equation}
\rho = \frac{1- \epsilon}{N}{\openone}_N + \epsilon \arrowvert 
\Psi \rangle \langle \Psi \arrowvert \;,
\end{equation}
where ${\openone}_N$ is the identity matrix of dimension $N$.
(In the high $\epsilon$ regime, an exact description can be obtained by 
reverting to the full Boltzmann distribution.) Since, for experimentally 
feasible values of $\epsilon$ (now and in the foreseeable future), the 
pseudo-pure state description suffices, we henceforth restrict our 
discussion to them.

Let us now consider quantum searching on pseudo-pure quantum states.
After $k$ iterations of the Grover search operator, one obtains the state
\begin{equation}
\rho_k = \frac{1- \epsilon}{N} {\openone}_N+ \epsilon \arrowvert 
\Psi_k \rangle \langle \Psi_k \arrowvert \label{pp} \;,
\end{equation}
where $\arrowvert \Psi_k \rangle$ is given by Eq.~(\ref{Psik})

Let us first look at the diagonal form of the reduced state of any remaining 
qubit (denoted by the index $\ell$) in the pure state given in 
Eq.~(\ref{Blochvec}). This has positive eigenvalues $\lambda_1(k)$, 
$\lambda_2(k)$ 
which are independent of $\ell$ but depend on the iteration index
$k$. These eigenvalues sum to one and their 
product, after some calculation, is given by
\begin{equation}
\lambda_1(k) \lambda_2(k)= {N(N-2) \over 2 (N-1)^2}\;
\sin^2 (2 k \theta_0)\cos^2 \theta_k \;.
\end{equation}
This allows us to decompose the {\it full\/} state at step $k$ into a 
Schmidt basis
\begin{equation}
\arrowvert \Psi_k \rangle = 
\sqrt{\lambda_1(k)} \arrowvert g' \rangle  \arrowvert  e \rangle 
- \sqrt{\lambda_2(k)} \arrowvert e' \rangle  \arrowvert  g \rangle \;,
\end{equation}
where $\{\arrowvert e \rangle, \arrowvert g \rangle\}$ describes an 
orthonormal basis for the $\ell$th qubit and 
$\{\arrowvert e' \rangle, \arrowvert g' \rangle\}$ is a pair of Hilbert 
space vectors for the other $(n-1)$ qubits. This expression will help us 
to derive a bound on the purity parameter.

In order to study the entanglement present in the pseudo-pure state
given by Eq.~(\ref{pp}) let us project it onto the 4-dimensional subspace 
spanned by the set $\{\arrowvert g'\rangle \arrowvert g\rangle$,
$\arrowvert g'\rangle \arrowvert e \rangle$,
$\arrowvert e' \rangle \arrowvert g \rangle$,
$\arrowvert e' \rangle \arrowvert e \rangle \}$.
The resulting 4-dimensional density matrix is
\begin{equation}
\rho_k^{(4)} = {N\over 4+\epsilon(N-4)} \Bigl(
{ 1- \epsilon\over N} \, {\openone}_4
+ \epsilon \arrowvert \Psi_k \rangle \langle \Psi_k \arrowvert \Bigr)
\label{4Dstate} \;.
\end{equation}
The boundary between separability and entanglement for such 4-dimensional 
states may be specified in terms of the so-called fidelity $F$.
The states~(\ref{4Dstate}) are separable when
$F\equiv\langle \Psi^- \arrowvert \rho_k^{(4)} \arrowvert \Psi^- \rangle 
\le \frac{1}{2}$ and are entangled otherwise \cite{bet}, where 
$\arrowvert\Psi^-\rangle\propto\arrowvert g'\rangle \arrowvert e\rangle 
-\arrowvert e'\rangle \arrowvert g \rangle$.
Since projection onto a subspace cannot create entanglement, it follows 
from this condition that if the original unprojected state~(\ref{pp}) 
is separable, then we must have
\begin{equation}
\epsilon \le \epsilon_k \equiv {1 \over 1+ N\sqrt{\lambda_1(k) \lambda_2(k)} }
\label{bound} \;.
\end{equation}
Similarly, the original state (at stage $k$ of the search) is entangled 
whenever $\epsilon > \epsilon_k$. Similar bounds have been earlier derived
when the pure state part was a maximally entangled state \cite{sam}.
However, the present bound quantifies the separability region of pseudo-pure 
states for each iteration $k$ during Grover's algorithm. Thus one can 
tell at each stage of computation whether the state is entangled for a 
given value of purity parameter.

\section{Speed-up and Entanglement for Pseudo-pure states}

Let us now return to a careful re-evaluation of what it means for a 
pseudo-pure state implementation of Grover's algorithm to demonstrate 
a speed-up compared to a classical search on a conventional computer. 
First, we note that the important common element in either a quantum or 
a classical search algorithm is the repeated need of evaluating the 
function $f$. Thus, we shall restrict our comparison to its number of 
evaluations. This is formally known as studying the {\it query 
complexity\/} of the problem. Second, since Grover's algorithm is 
probabilistic, there is no upper-bound to the worst-case number of 
evaluations.   Instead, we study the expected number 
of evaluations for finding the target. To be fair then, we must compare 
this with the expected number of function-evaluations on a conventional 
computer. If we exclude the use of an exponentially growing auxiliary 
memory to store failed trials then this classical query complexity is 
\begin{equation}
{\cal N}_{\rm class}=\frac{(N+2)(N-1)}{N}
\end{equation}
for finding a single object out of 
$N$ entries. This value may be achieved by systematically stepping
through the $N$ possible locations for the object and evaluating $f$
to see if the object is there. If it is not found in this way by step 
$N-1$ then we would know it is at the final location $N$. (The specific 
value for classical query complexity quoted in Ref.~\cite{vetal} 
for $N=8$ can be obtained from our general result ${\cal N}_{\rm class}$
above).

How does this compare to the expected number of function evaluations 
for quantum search on pseudo-pure state implementations?
The probability of finding the target state after $k$ iterations is 
\begin{equation}
p(k) = \langle y \arrowvert  \rho_k
\arrowvert y \rangle= \frac{[1 + \epsilon (N \sin^2 \theta_k - 1 )]}{N}<1.
\end{equation}
This probability must be amplified statistically through repetitions 
or parallelism.
The expected number of repetitions required to 
identify the target is just the reciprocal of this probability. Each 
such repetition involves $k$ function-evaluations for each run of the 
algorithm, plus one to test the result. In order to give Grover's 
algorithm its maximal advantage, we shall optimize the speed (rather
than the probability). Thus, the optimal expected number of 
function evaluations is just 
\begin{equation}
{\cal N}_{\rm pseudo}= \frac{\min_k (k+1)}{p(k)}
\equiv \frac{(k_{\rm opt}+1)}{p(k_{\rm opt})},
\end{equation}
where $k_{\rm opt}$ is the optimal number of iterations of the Grover
operator. Thus, a pseudo-pure state quantum computer can search faster 
than a conventional computer, provided
\begin{equation}
{\cal N}_{\rm pseudo}< {\cal N}_{\rm class} \label{opt} \;.
\end{equation}

In the above analysis we have utilized the standard projective measurement
which is a desirable operation in any quantum information processing unit.
However, the liquid-state NMR machines use weak measurements 
and the success probability obtained in a strong measurement
is always greater than the success probability for a weak 
measurement. So in our analysis we give extra benefit to the experimental
situation in tolerating the error. In any case, if we take weak measurement
the situation for NMR machines will be no better (as that will increase the
number ${\cal N}_{\rm pseudo}$).

We have now developed the tools to effectively answer the question:
{\it Can Grover's algorithm yield a speed-up on any existing pseudo-pure
state implementations\/} \cite{ch,jmh,vetal0,vetal}? Since all liquid-state 
NMR experiments performed so far have generated only separable ensembles
\cite{sam}, with purity parameters that fit well our pseudo-pure state
treatment above, let us see what effect separability has on efficiency.
From Eq.~(\ref{bound}) separability throughout the search implies 
$\epsilon \le \min_{k=0}^{k_{opt}}\,\epsilon_k$. Thus, separability
places an upper bound on the purity parameter (see values for this
bound in Ref.~\cite{sam}) and hence a lower bound, 
${\cal N}_{\rm pseudo}^{\rm (min)}$, on the quantum query complexity. 
This lower bound is given in Table~1 for the optimized Grover algorithm 
for search spaces up to $n=8$ qubits. (The trend we find continues for
arbitrarily large numbers of qubits.) The surprising observation is that, 
for more than $4$ qubits, the optimal quantum strategy is not to use the 
liquid-state NMR machine at all and simply guess the answer (equivalent to
sampling the initial random state). Clearly, this is never as good as a 
systematic conventional search.

{$~$}

Table 1
$~~~~~$
\begin{tabular}{crcrr}
$n$&~~$~~N$&~~ $k_{\rm opt}$~ & ${\cal N}_{\rm pseudo}^{\rm (min)}\!\!$
&~~ ~${\cal N}_{\rm class}\!\!$\\

    1 & 2        & 0      & 2\,~~~    & 1~~~~~\\
     2 & 4        & 1      & 2\,~~~    & 2.25\\
   3 & 8        & 1      & 5.48      & 4.38\\
     4 & 16       & 2      & 12.89     & 8.44\\
  5 & 32       & 0     & 32\,~~~    & 16.47\\
    6 & 64       & 0     & 64\,~~~    & 32.48\\
     7 & 128      & 0    & 128\,~~~    & 64.49\\
     8 & 256      & 0    & 256\,~~~    & 128.50\\
\end{tabular}

{$~$}

Curiously, we note that our table shows a speed-up even without 
entanglement for the $n=2$ qubit implementation. That this algorithm 
requires no entanglement in this case, has also been noted for the 
pure-state implementation of Grover's algorithm \cite{lloyd} (with a 
similar result for the two qubit implementation of the Deutsch-Jozsa 
algorithm \cite{collin}). This occurs because the separable target state 
can be reached by a {\it single\/} application of the Grover operator from 
the separable initial state; hence the evolution has passed through 
no entangled states. Without entanglement in the two-qubit case there 
is also no penalty to the efficiency when enforcing separability through 
a reduced purity parameter. Despite this curiosity, this speed-up is only 
possible for unphysically large purity parameters, 
$\epsilon > 23/27 \simeq 0.852$, where the pseudo-pure description is no 
longer valid for liquid-state implementations. This observation and the 
above analysis show that {\it entanglement is necessary for obtaining 
a speed-up in Grover's algorithm on a liquid-state NMR machine relative 
to a classical computer}.  Let us emphasize that for pure-state machines 
(e.g., ion-traps) the two-qubit implementation of Grover's algorithm 
is realistic and will yield a speed-up of 2.25 (omitting the now superfluous 
final testing function-evaluation).

As we have seen, the presence of some entanglement is essential to obtain 
a speed-up, except in the two-qubit case. But how much of it?  Let us 
impose the speed-up condition~(\ref{opt}) to obtain a lower bound 
$\epsilon_{\rm speed-up}(k_{\rm opt})$ on the purity parameter within the
pseudo-pure state approximation. After iteration $k$, condition~(\ref{bound}) 
implies entanglement is present whenever 
$\epsilon_{\rm speed-up}(k_{\rm opt})>\epsilon_k$. We studied this relation 
numerically for $2< n\le 20$ qubits with 
$0<k\le k_{\rm opt}\le \lceil\pi/4\theta_0 - 1/2\rceil$, and found 
that entanglement was present after every iteration except possibly 
the last step of the algorithm when $\theta_{k_{\rm opt}}>\pi/2$. 
Thus, we may draw the much stronger conclusion that, for more than
two qubits, {\it some degree of entanglement\/} is necessary during 
the {\it entire\/} computation in order to obtain any speed-up for
Grover's algorithm on a liquid-state NMR machine.


To better understand the repercussions of this result, let us directly 
address the main objections raised about the role of entanglement in 
performing real quantum computation in this system. The most dismissive 
argument has been that quantum computational efficiency derives {\it only\/} 
from the unitary evolution of quantum states, but is independent of the 
type of states being used (Laflamme in Ref.~\cite{PT}). Now, for the 
sake of clarity, let us adopt the definition of quantum computing suggested 
by the same proponent in the same context, namely: Quantum computing consists 
of efficiently evolving from initial to final density matrices by unitary 
operations. Since it is not enough to reach the final state 
probabilistically, we have calculated the query complexity of the problem 
in detail (assuming perfect unitary evolution). Thus our analysis clearly 
proves that unitarity alone is not enough to achieve quantum efficiencies.

A different objection correctly points out that liquid-state NMR
machines are not exactly described by the pseudo-pure state formalism, and
hence the bounds we derive above for entanglement and separability may not
be applicable.  Indeed, corrections to this bound are necessary in the
few-qubit regime, where the bounds for the purity parameter are high.
However, in most likelihood, these corrections will not alter our results
and may even strengthen them (by raising the lower bound for entanglement).
Further, noting that existing liquid-state NMR machines are remarkably
far from reaching this range of purity parameters, any possibility of
speed-up is quite out of the question.

\section{Dynamics of Pseudo-pure versus pure states}

Granted that entanglement is needed to achieve quantum efficiencies,
it has been argued \cite{lp} that the ability of liquid-state NMR 
machines to physically follow unitary quantum evolution qualifies them as
efficient simulators (without speed-up). The reasoning is based on the 
realization that the observable operators are traceless \cite{ww}. Thus, 
the expectation value of any traceless observable on a pure quantum state is 
the same, up to a scale factor, as it would be on a pseudo-pure state. 
This is because for a traceless observable $\Theta$ the average
in pseudo-pure state is 
${\rm tr}\,(\rho \Theta) = \epsilon \langle \Psi |\Theta| \Psi \rangle$. 
However, in the quantum world fluctuations play as important 
a role as expectation values. In general, if an observable is traceless its 
square need not be. In this case the quantum fluctuations of an operator 
on a pure state are {\it not} equal to those fluctuations on the 
corresponding pseudo-pure state. For example, if $\Theta$ is a traceless 
operator its root-mean-square fluctuations on pseudo-pure states are 
determined by
\begin{equation}
(\Delta \Theta)_{\rm pseudo}^2 = \epsilon (\Delta \Theta)_{\rm pure}^2
+ (1-\epsilon) \Bigl(\frac{{\rm tr}\,(\Theta^2)}{N} + \epsilon
\langle \Theta \rangle_{\rm pure}^2 \Bigr) \;.
\end{equation}

As a simple example, consider 
$\Theta = d \rho/d \epsilon=P-\openone /N$, where 
$P = \arrowvert \Psi \rangle \langle \Psi \arrowvert$. It is a traceless 
operator, but its square is not. The uncertainty of the operator $\Theta$ 
on the pure state is zero, whereas on the pseudo-pure state it is
\begin{equation}
(\Delta \Theta)_{\rm pseudo}^2 = (1 - \epsilon)
\Bigl(1- \frac{1}{N}\Bigr)\left[ \frac{1}{N} 
- \epsilon \Bigl(1- \frac{1}{N}\Bigr) \right] \;.
\end{equation}
In fact, very rarely will it be the case that all the moments (even for 
the restricted class of traceless operators) are related by simple 
scale-factors to those produced by a liquid-state NMR quantum computation.
Thus, despite the correct expectation values being accessible for traceless 
operators, {\it we cannot say that liquid-state NMR quantum computers have 
good dynamics}.

\section{Conclusion and discussion}

In conclusion we have shown that the original version of Grover's
algorithm implemented on qubits necessarily generates quantum entanglement
during the computation process. Further, we have shown that a quantum 
computer in a non-entangled pseudo-pure state requires more iterations 
than a classical computer to perform a virtual database search if one 
uses our method of counting the number of queries. Thus, for any existing 
liquid-state NMR set-ups our results preclude any possibility of speed-up 
in running Grover's algorithm. This conclusion is based on exact 
calculations for pseudo-pure state implementations, which show that 
entanglement is essential {\it throughout\/} the computation. Moreover, 
even a modest reduction of entanglement is tantamount to a total loss 
of speed-up. Despite the decisive results reached for pseudo-pure state 
implementations of this algorithm, we found that entanglement-free 
speed-up would be possible for the two-qubit case (search space of size 
four) of Grover's algorithm on pure-state machines such as ion-trap 
quantum computers. Our analysis has been with respect to a specific
non-asymptotic regime and does not find a way for scalable computation 
without entanglement. Finally, our analysis does not resolve how other 
NMR experiments should to be interpreted such as quantum teleportation
without entanglement.

{\it Note added:} After we completed this work, a 
recent article has even implemented the quantum search algorithm using only 
the wave nature of classical Fourier optics \cite{bhata} which involves
{\it no\/} entanglement! However, one should not be surprised into thinking
that this is a contradiction to our findings presented here. What happens 
with classical devices implementing the search algorithm is that 
the number of resources needed increases prohibitively with the size of 
the input.

{$~$}

{\bf Acknowledgements:} SLB thanks N.\ Cohen for discussions. AKP gratefully 
thanks EPSRC for financial support for the year 1998-2000 during which 
this work was carried out.


\end{document}